\documentclass[net-ND]{svjour}
\usepackage[tbtags]{amsmath}
\usepackage{amssymb}
\usepackage{graphicx}
\usepackage{txfonts}
\usepackage{epstopdf}

\begin{document}

\title{IMPACT OF THE PHONON COUPLING ON THE DIPOLE
STRENGTH AND RADIATIVE NEUTRON CAPTURE}

\titlerunning{Impact of the phonon coupling on ... radiative neutron capture.}

\author{ A. AVDEENKOV \inst{1,} \inst{2}
   \and S. GORIELY \inst{3}
   \and S. KAMERDZHIEV \inst{4} \cormail{kamerdzhiev@ippe.ru} }

\authorrunning{A.Avdeenkov, S. Goriely, S. Kamerdzhiev,}

\institute{ \and National Institute for Theoretical Physics , Stellenbosch
Institute of Advanced Study, 7602 South Africa
\and Institute of Nuclear Physics, Moscow State University , Moscow 119992, Russia
   \and Institut d'Astronomie et d'Astrophysique, ULB, CP 226, B-1050 Brussels, Belgium
   \and Institute of Physics and Power Engineering, 249033, Obninsk, Russia
}

\abstract{The E1 strength functions and radiative capture cross sections for several compound Sn isotopes, including unstable
$^{132}$Sn and $^{150}$Sn, have been calculated using the self-consistent microscopic theory. In addition to
the standard RPA or QRPA approaches, the method includes the quasiparticle-phonon coupling and the single-particle continuum. The results
obtained show that the phonon contribution significantly affects the pygmy dipole resonance, which is of particular relevance for a proper description of the radiative neutron capture. The impact of the phonon coupling on the pygmy dipole resonance and the radiative
neutron capture cross sections increases with the ($N-Z$) difference. For example, in the (0-10) MeV interval the full
theory gives $17\%$ of the energy-weighted sum rule for $^{150}$Sn and $2.8\%$ for $^{124}$Sn, whereas the continuum QRPA approach gives $5.1\%$
and $1.7\%$, respectively. These results indicate the importance the self-consistent calculation can have, especially when applied to neutron-rich nuclei of  astrophysical interest. The comparison with the  widely-used phenomenological Generalized Lorentzian approach shows  that the (Q)RPA approach gives an increase in  the neutron capture cross section by a factor of 2 for
$^{132}$Sn and a factor of 10 for $^{150}$Sn and  that the inclusion of the phonon coupling still  increases these cross sections even furhter, by a factor of 2--3.
}


\maketitle

\section{Introduction}
Inclusion of the coupling of single-particle degrees
of freedom with phonon degrees (in short, phonon coupling or PC), 
in addition to the standard  Random Phase Approximation (RPA) for magic 
nuclei or the Quasi-particle RPA (QRPA) for non-magic nuclei, 
is a main line of the recent development of microscopic
nuclear theory. In particular, the Extended Theory of Finite Fermi Systems
(ETFFS) \cite{refn1} includes both PC and the single-particle continuum,
 the latter being absolutely necessary for nuclei with the nucleon separation
 energy near zero. This approach has been recently generalized to include
 pairing in the quasiparticle time-blocking approximation
 \cite{refn2}, (ETFFS(QTBA)).
 This and other  approaches developed with the PC have been supplemented
with consideration of the self-consistency between
the mean field and the effective interaction \cite{refn3,refn4,refn5}.
On this basis, it was possible to perform a transition from the two
sets of parameters used (the first one for the effective force
and the second one for the mean field) to one unique set based on 
Skyrme force.
 Both  these improvements --the single-particle continuum and  the self-consistency--
are of great  importance, for
astrophysics as well as for   nuclear data evaluations, since
they provide a reliable framework to calculate
the structure of exotic nuclei, especially those with great
neutron excess and/or with the nucleon separation energy
close to zero.

The PC role in the description of giant resonances in stable
nuclei is well known: it explains approximately $50\%$ of
the observed widths, their gross structure and sometimes even the
fine structure \cite{refn1}. However, the direct influence of  the PC on
giant resonances in unstable nuclei is less studied.
It is clear that the PC role should be important for the
widths here as well. But so far, the characteristics of the giant resonances in
unstable nuclei have only been systematically studied within the (Q)RPA
 \cite{refn6,refn7,refn8}   approach.

The PC plays an important role in the determination of the so-called pygmy dipole resonance
(PDR), which is the low-energy part of the giant dipole  resonance (GDR)
that can  exhaust about $1-2\%$ of the  energy-weighted sum rule (EWSR). 
At present there is no clear understanding around  some important questions
related to PDRs, see \cite{refn6,refn9}.
This resonance is also known to be of particular relevance in the description 
of the radiative neutron capture \cite{refn8,refn10}. However, the PDR has 
essentially been treated on the basis of phenomenological models up to now.
The question arises for exotic nuclei where the
phenomenological approach may fail  because of the
specific features of the PDR in such nuclei \cite{refn6,refn8}.
The answer is the same as for the problem of giant resonances
in  unstable nuclei and is quite obvious: it is necessary to
use a reliable theory for such nuclei, i. e., as discussed
above, a self-consistent theory which accounts for the PC and
the single-particle continuum in addition to the standard
(Q)RPA.

In our previous works \cite{refn4,refn9,refn11} we  developed the self-consistent
version of the ETFFS(QTBA) using a discretization
procedure for the single-particle continuum with different
kinds of Skyrme forces, including SLy4, where
the  velocity-dependent parts of forces were considered in the local approximation
(we called it DTBA).  The consequence of such a simplification is the necessity to
renormalize the interaction in order to obtain the
spurious state at zero energy.

The main goals of the present work  are: i) to study the PDR and GDR consistently in
stable $^{124}$Sn and unstable $^{132}$Sn, $^{150}$Sn isotopes using the variant (DTBA) of the
microscopic self-consistent version of the ETFFS;
ii) to investigate the impact of  the PC on the radiative
neutron capture cross section.

\section{ Self-consistent calculations of the PDR and GDR }
We use  the SLy4  Skyrme
force \cite{refn12} which proves to be rather successful
in description of ground states and some excited
states within  the (Q)RPA. The ground states are calculated
within the   Hartree-Fock-Bogolyubov approach using the HFBRAD code \cite{refn13}.
The residual interaction for the (Q)RPA and the
following QTBA calculations is derived as the second
derivative of the Skyrme functional (see details
 in \cite{refn4}).

 Fig.~1 shows the photoabsorption cross sections
for $^{124}$Sn and $^{150}$Sn nuclei. A more detailed discussion on $^{132}$Sn can be found
in \cite{refn11}.
One can see that the phonon contribution
is noticeable for  the  PDR and it is
increased with the (N-Z) difference growth. For example, in
the (0-10) MeV interval the full theory gives $17\%$ of EWSR
for $^{150}$Sn and $2.8\%$ for $^{124}$Sn, whereas within the
QRPA approach we have $5.1\%$ and $1.7\%$, respectively.

\begin{figure}[ht]
\centering
\resizebox{0.90\columnwidth}{!}{
   \includegraphics{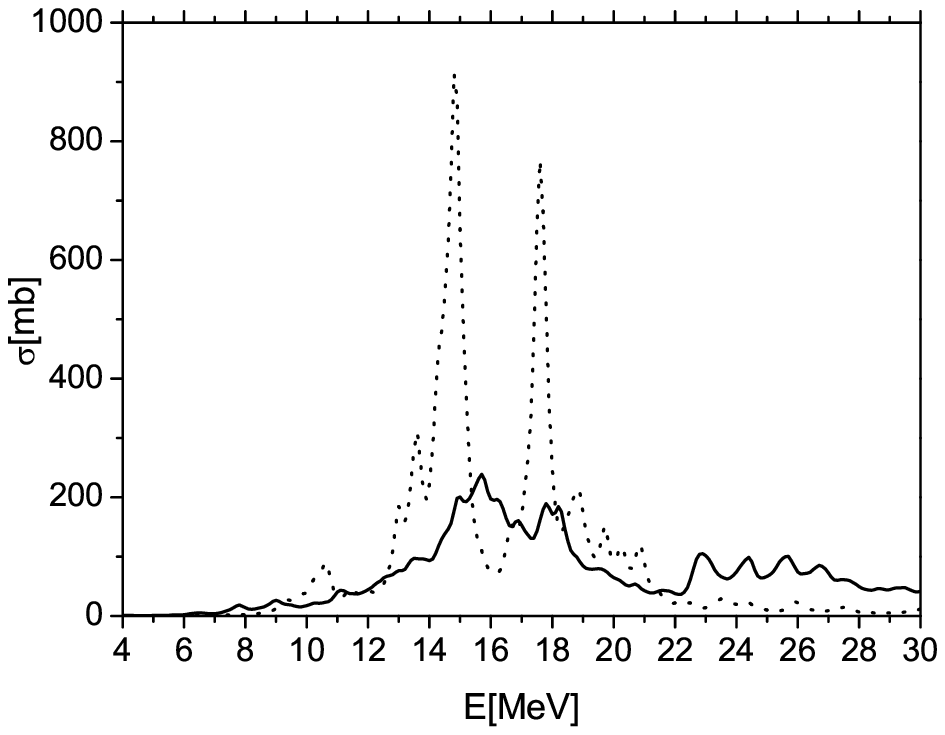}
}
\resizebox{0.90\columnwidth}{!}{
   \includegraphics{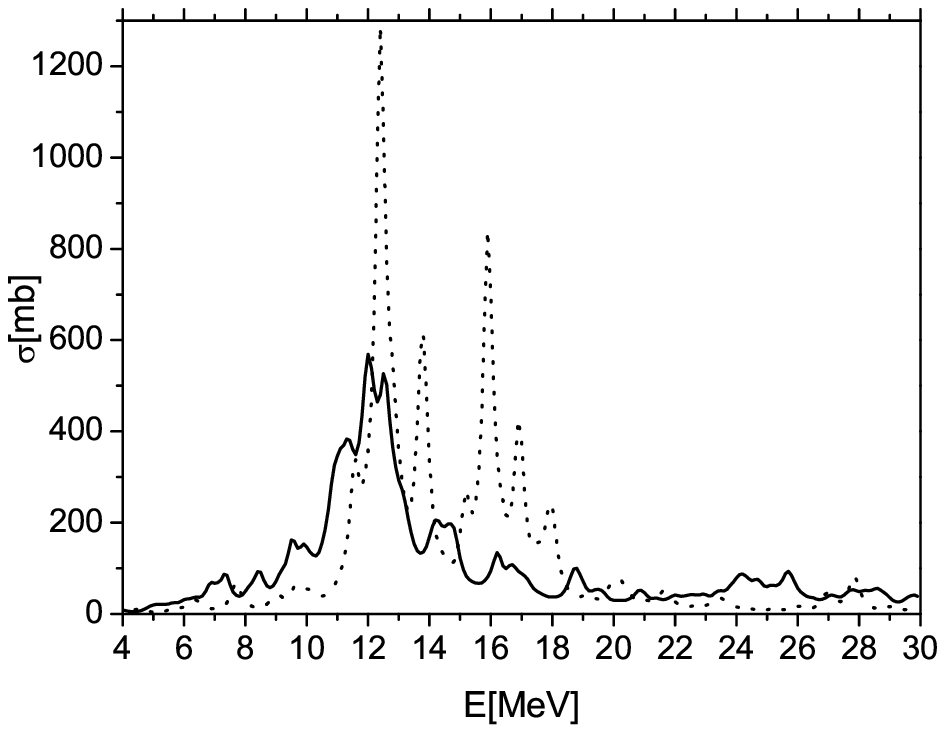}
}
\caption{Photoabsorption cross sections for $^{124}$Sn (top) and $^{150}$Sn
(below) calculated within the self-consistent ETFFS(DTBA)
without (dotted curves) and with (solid curves) the PC.}
\label{fig:1}
\end{figure}

 Fig.~2 shows the comparison of the E1 strength functions obtained
in the framework of our self-consistent DTBA version with the
(Q)RPA calculations. We used a normalization procedure
here. Namely, the DTBA strength has been folded with a
Lorentzian of 1 MeV width for DTBA and 2.7 MeV for
RPA in order to  broaden the GDR to a typical width of
4 MeV, as observed experimentally for the stable nuclei
 Although the DTBA peak energy of $^{132}$Sn is
predicted at a relatively low energy (13.5 MeV) with respect
to experimental data (16.1 MeV), the PDR is correctly
reproduced around 9.8 MeV \cite{refn14}.

In Fig.~2,  our results are also compared with those obtained with the 
phenomenological Generalized Lorentzian strength function \cite{refn15}.
The Lorentzian approach is widely used for practical application, though 
it suffers from shortcomings of various sorts. On the one
hand, the location of the GDR maximum energy and width remain to be predicted from some
underlying model for each nucleus. For many applications, these properties have
often been obtained from a droplet-type of model or from experimental systematics  \cite{refn10}.
In addition, the Lorentzian model is unable to predict the  $E1$ strength at
 energies below the neutron separation energy. Different parametrizations or functional forms (including in particular an energy- and temperature-dependent width) have been proposed (see e.g. \cite{refn15}) to reconcile experimental data in the photon or radiative neutron capture channels, but none of the proposed closed forms can nowadays explain the various trends observed at low energies. Most of all the Lorentzian approach cannot provide any predictions on the low-energy PDR, neither on its presence, nor on its characteristics. 
For this reason, it is of particular interest to analyze to what extent our predictions based on self-consistent microscopic models differ from those used in practical applications. 
One can see that the difference between
\cite{refn15}, on the one hand, and both the QRPA and DTBA, on the other hand,
 increases  with the ($N-Z$) difference.  As our preliminary calculations
have shown, the main reason
is that the $A$-dependence of  the integral
characteristics of neutron-rich nuclei differs strongly from the usual ones.
This fact demonstrates
the non-applicability of the standard systematics~(e.g. $E_0=80A^{-1/3}$)
for unstable nuclei.

\section{Calculations of radiative neutron capture cross sections}

\begin{figure}[ht]
\centering
\resizebox{0.90\columnwidth}{!}{
   \includegraphics{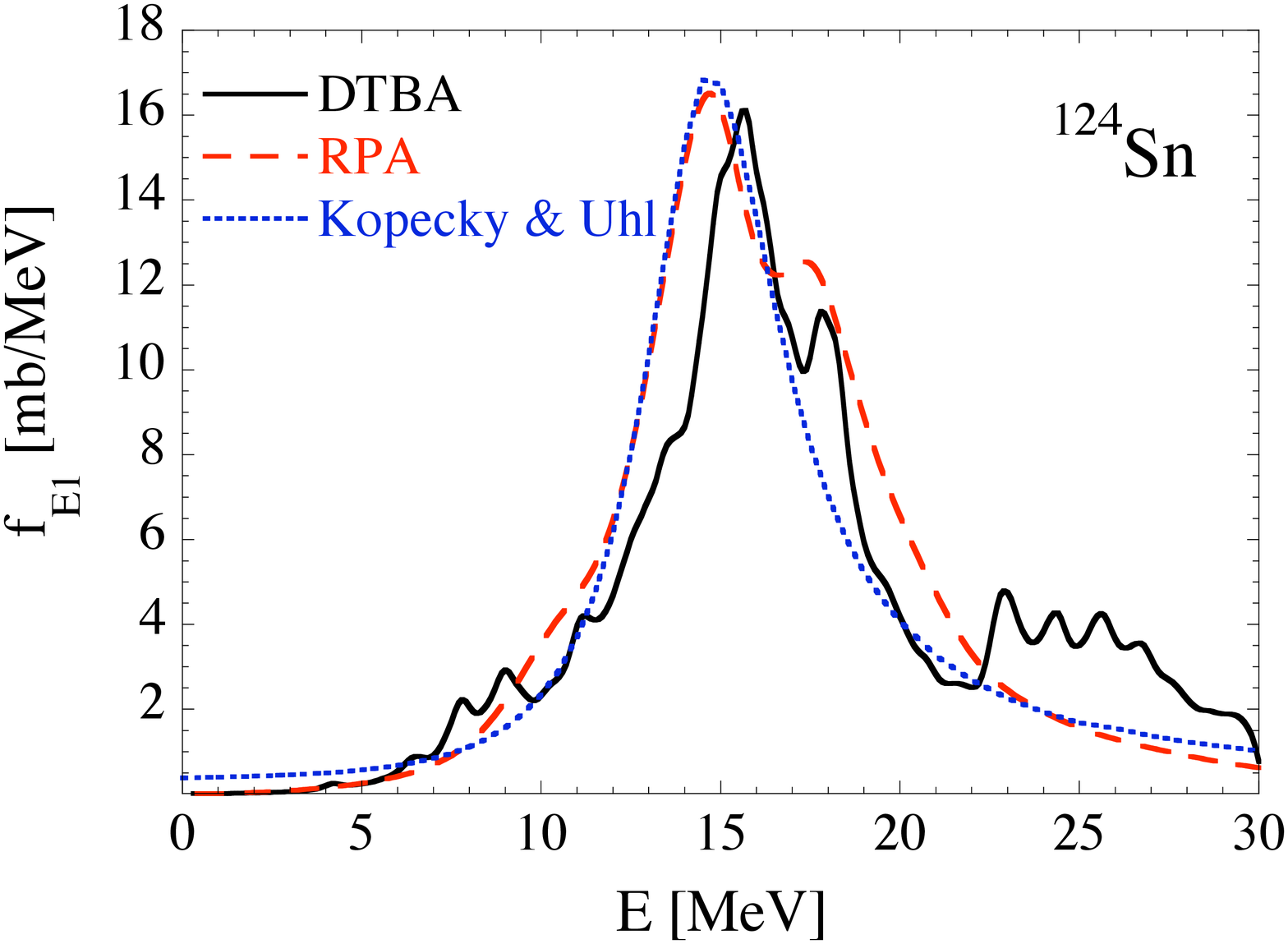}
}
\resizebox{0.90\columnwidth}{!}{
   \includegraphics{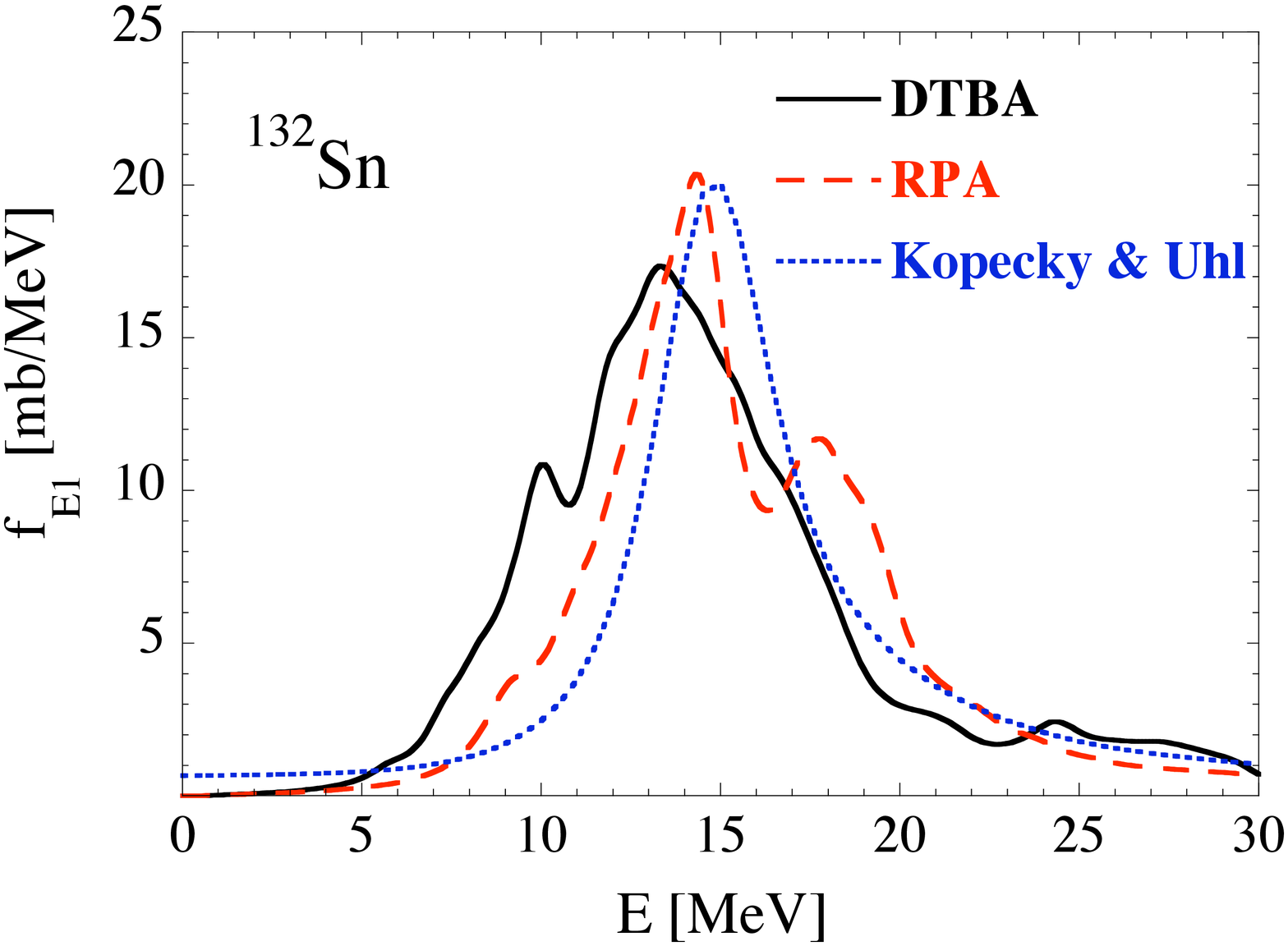}
}
\resizebox{0.90\columnwidth}{!}{
   \includegraphics{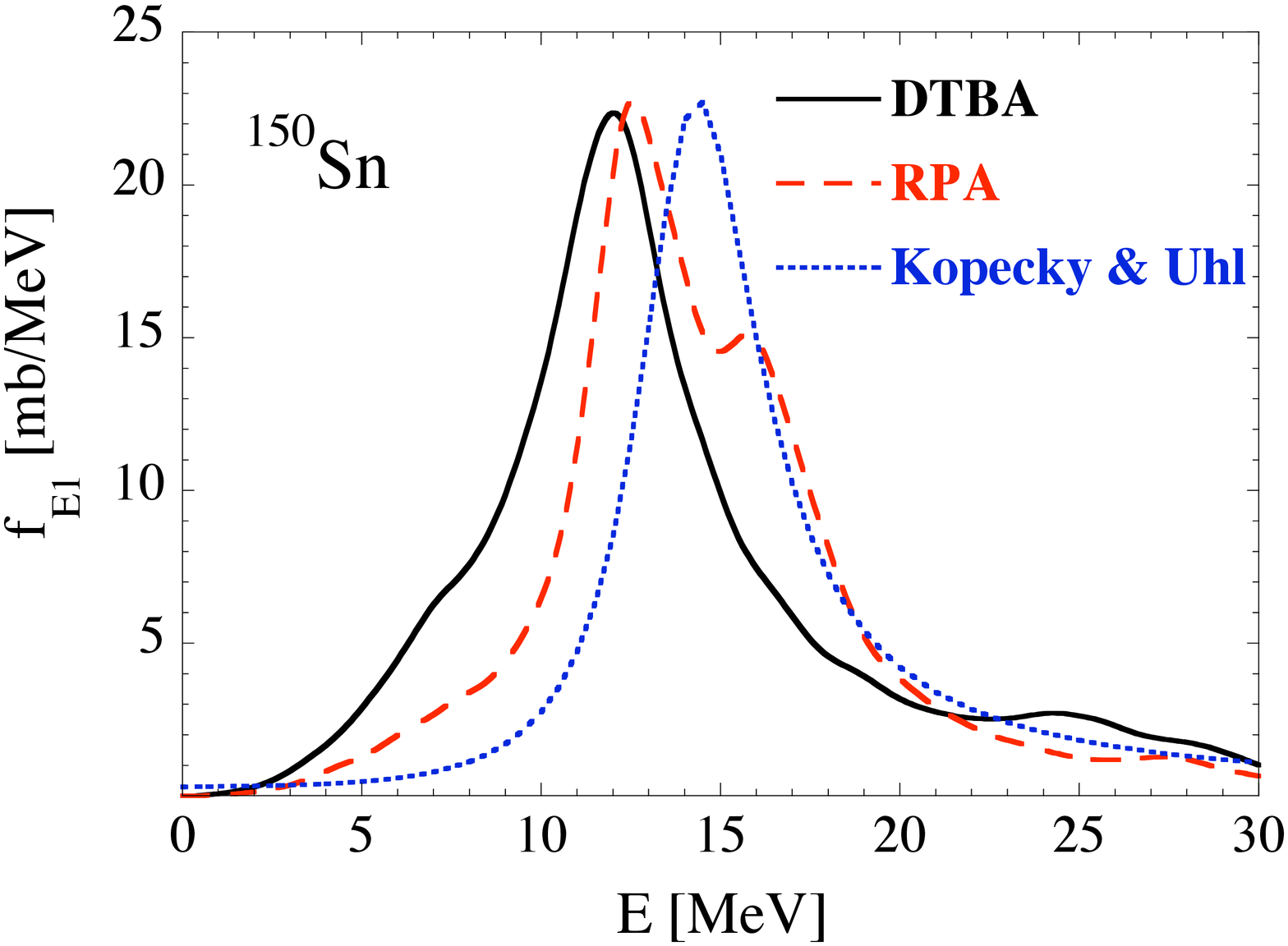}
}
\caption{Strength functions for the isovector 1$^-$ states
in $^{124}$Sn, $^{132}$Sn and $^{150}$Sn nuclei.
The solid, dashed and dotted curves are for the DTBA, (Q)RPA and  Kopecky-Uhl \cite{refn10,refn15} approaches,
respectively}
\label{fig:2}
\end{figure}

\begin{figure}[ht]
\centering
\resizebox{0.90\columnwidth}{!}{
   \includegraphics{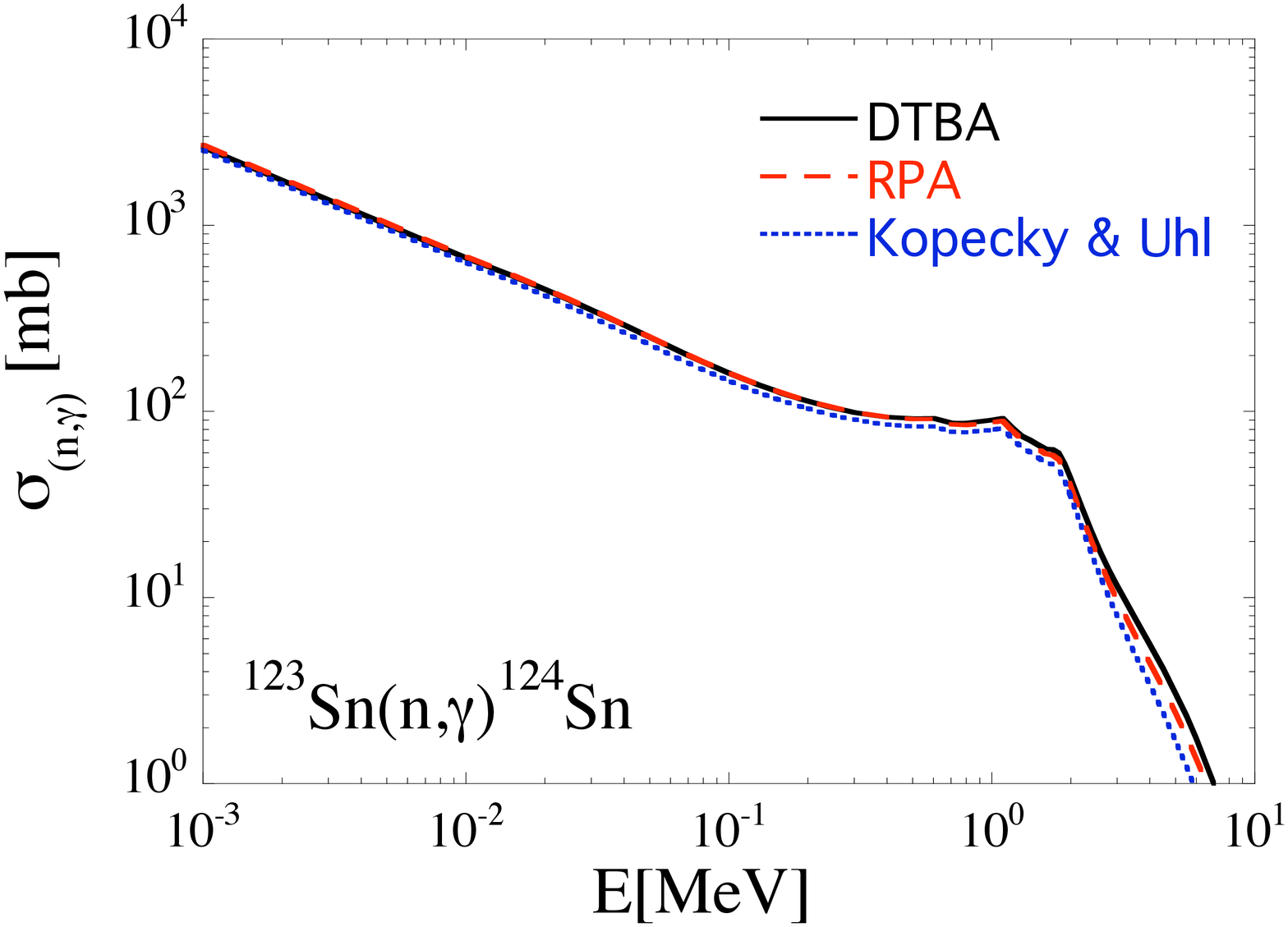}
}
\resizebox{0.90\columnwidth}{!}{
   \includegraphics{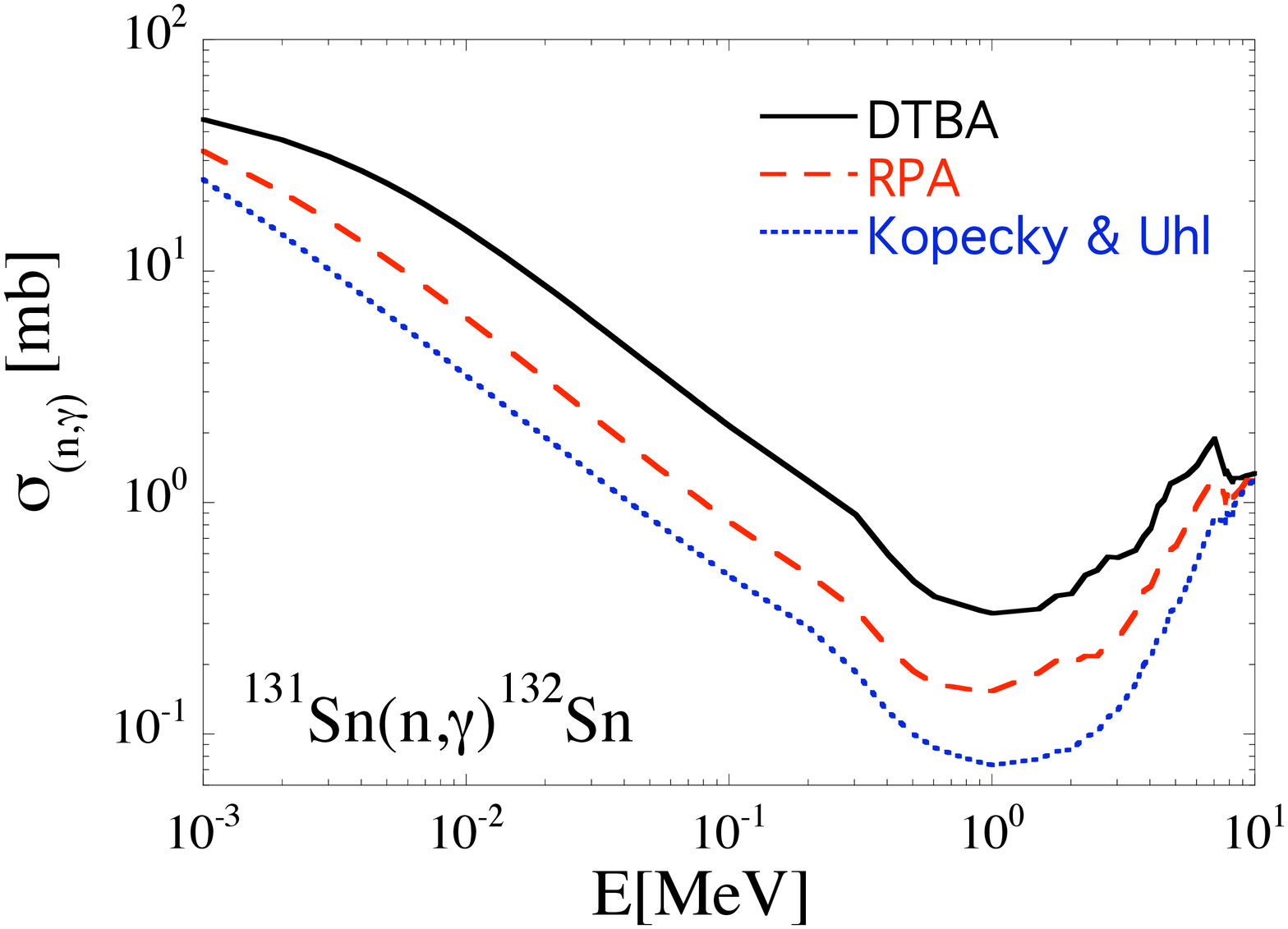}
}
\resizebox{0.90\columnwidth}{!}{
   \includegraphics{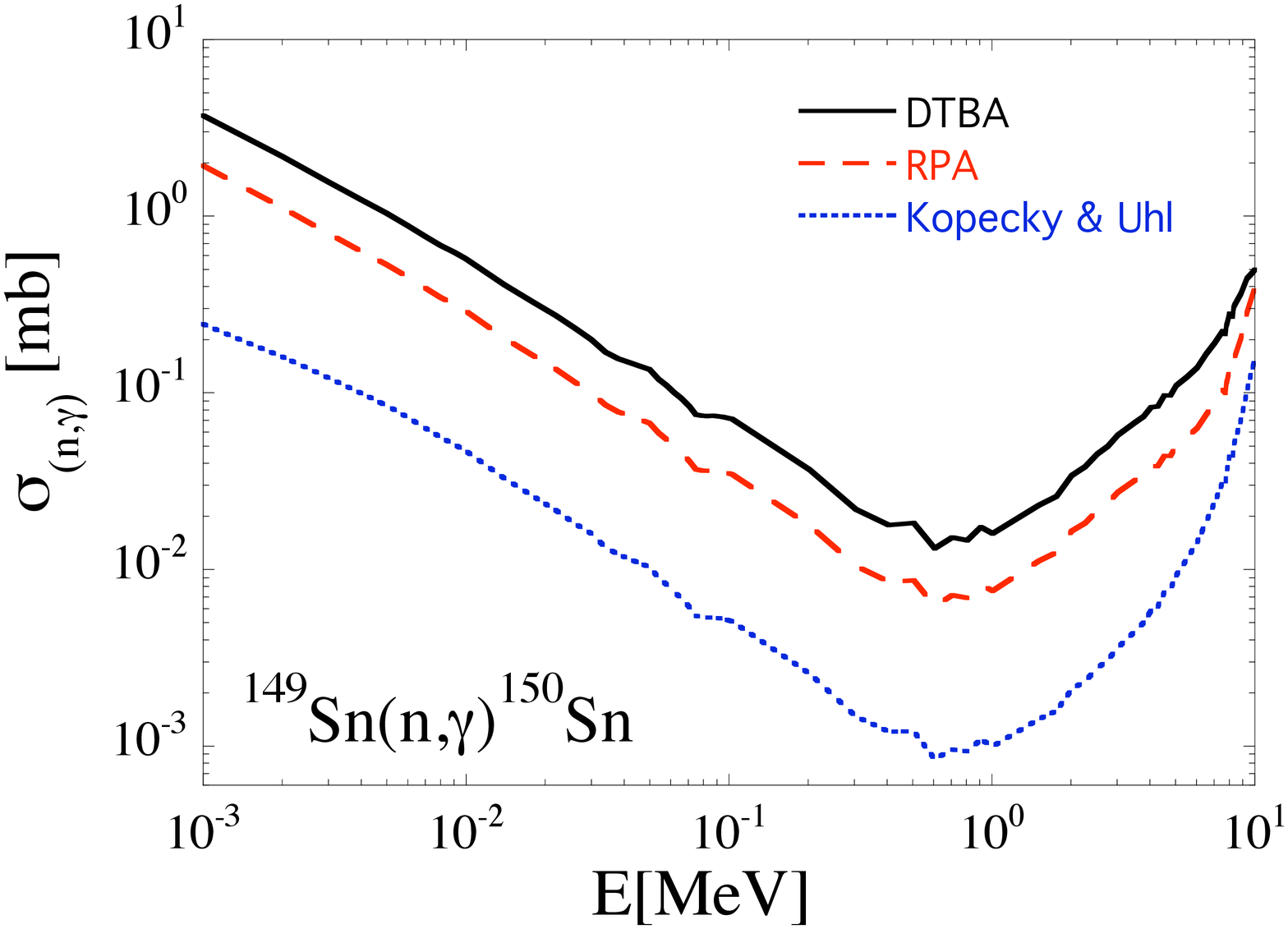}
}
\caption{Radiative neutron capture cross sections  obtained with the $\gamma$-ray strength functions
of $^{124}$Sn, $^{132}$Sn, $^{150}$Sn shown in Fig.~2}
\label{fig:3}
\end{figure}

The radiative neutron capture cross sections have been
calculated using the reaction code TALYS \cite{refn16}.
The DTBA
strength functions with and without PC have been included
in the calculation of the electromagnetic de-excitation transmission
coefficients. The final radiative neutron capture
cross sections calculated with the strength functions of Fig.~2 are shown in Fig.~3. We show the cross
sections leading to the $^{124}$Sn, $^{132}$Sn and $^{150}$Sn compound
nuclei, respectively. Note that such reactions are of interest
for the r-process nucleosynthesis, and illustrate the impact that the low-lying
strength can have on cross sections, and consequently on
reaction rates. As seen in Fig.~3,  including  the PC
increases the cross section by a factor of 2--3. In
all   cases the trend is clear: 
 the cross sections  follow on the whole
the strength functions shown in Fig.~2, and the DTBA with
 an  extra low-lying strength gives a final reaction cross
section (and Maxwellian-averaged rate around $T = 1-2 \cdot 10^9K$)
 about 3 times larger  compared with the prediction
obtained with the HFB+QRPA calculation \cite{refn7}. For the case of 
the compound $^{124}$Sn, we see in Fig.2 that the difference between the DTBA and QRPA strength 
functions is much smaller than for $^{132}$Sn and $^{150}$Sn. It is necessary also to note here that 
the theoretical consideration for $^{124}$Sn is based in a sense on the known experiment, 
 this is especially true  
 for the Kopecky-Uhl approach (see below). For these reasons,  the appropriate cross sections 
in Fig.3 are insensitive by sight to the three strength functions.

In Fig.~3 we also compare  the cross sections with the widely used
Generalized Lorentzian strength function \cite{refn10,refn15}. We have found
that for the stable $^{124}$Sn nucleus the cross section is almost
identical to the one obtained with the ETFFS approach,
although the strength functions can differ significantly at
energies below the neutron separation energy. For
neutron-rich isotopes, the QRPA strength gives a significant
increase in the cross sections with respect  to those obtained with the Lorentzian function of  \cite{refn15}, namely by a factor of 2 for $^{132}$Sn and a factor of 10 for $^{150}$Sn. This
confirms the results of \cite{refn7}. Moreover, the inclusion
of  PC  increases the
cross sections for the unstable $^{132}$Sn and $^{150}$Sn by a factor
of 2--3 at the energies around 100 keV, as illustrated in Fig.~3.
These results  demonstrate the necessity  of self-consistent
approaches for  the calculation of the radiative capture cross sections of exotic neutron-rich nuclei.

\section{Conclusion}
The electric dipole strength function has been calculated
on the basis of the ETFF(DTBA) model which simultaneously
takes into account  the RPA or QRPA configurations,
the more complex 1p1h$\otimes$phonon
configurations and the single-particle continuum.
These are the necessary ingredients  for a consistent study  of the
giant and pygmy resonances.

 It was demonstrated that the DTBA approach predicts a relatively large low-lying strength compared to the
(Q)RPA  one around 10 MeV in the
neutron-rich Sn isotopes.
The radiative neutron capture cross sections were calculated
with the DTBA and (Q)RPA strength functions and
shown to be sensitive to the predicted low-lying strength.
In particular, the calculations  demonstrated that  including the PC 
leads to a significant increase in the reaction cross section.
The comparison with the  Generalized Lorentzian approach confirmed
the non-applicability of the phenomenological approaches for neutron-rich
nuclei.

The ETFFS(DTBA) approach is believed to provide
a more complete and coherent description of the gamma-ray
strength function than the previous models used so far. For
astrophysics applications in particular, such calculations are
highly recommended for a more reliable estimate of the
electromagnetic properties of exotic nuclei.

\begin{acknowledgement}
The work was partly supported by the DFG grant
No. 436RUS113/994/0-1 and RFBR grant No.09-02-
91352NNIOa
\end{acknowledgement}

\end{document}